\documentstyle{article} 
\newcommand{\be}{\begin{equation}}
\newcommand{\ee}{\end{equation}}
\newcommand{\bea}{\begin{eqnarray}}
\newcommand{\eea}{\end{eqnarray}}
\newcommand{\beax}{\begin{eqnarray*}}
\newcommand{\eeax}{\end{eqnarray*}}
\newcommand{\no}{\nonumber\\}
\def\r{{\bf r}}
\def\k{{\bf k}}
\def\p{{\bf p}}
\def\x{{\bf x}}
\def\y{{\bf y}}
\def\z{{\bf z}}

\begin{document}
\hfill  CTP \# 2514

\hfill  February, 1996

\vskip 0.5in                         
\centerline{\large Methods for a Nonuniform Bose Gas}
\vskip 0.25in  
\centerline{ Kerson Huang and Paolo Tommasini\footnote
{ Present address: Institute for Theoretical Atomic and Molecular Physics,
Harvard-Smithsonian Center for Astrophysics, Cambridge, MA 02138. 
} }
\vskip 0.25in
\centerline{Center for Theoretical Physics, Laboratory for 
Nuclear Science}
\centerline{       and Department of Physics}
\centerline{   Massachusetts Institute of Technology}
\centerline{   Cambridge, MA 02139}
\vskip 0.5in
\centerline{                ABSTRACT}
\bigskip
We review mathematical methods for the treatment of a system of Bose
particles with nonuniform density. Common pitfalls are pointed out.

\section{Introduction}

The observation of Bose-Einstein condensation in atomic traps
[1]-[3] has renewed theoretical interest in studying a system
of Bose particles with a nonuniform density. The experiments involve
atoms of mass $m$ condensed in a harmonic trap characterized by 
frequency $\omega$. The relevant parameters, as discussed more fully 
in [4] and [5], have the following orders of magnitude:
\bea
a &\sim& 10^{-6}{\rm\ cm},\no
\lambda &=& (2\pi\hbar^2/mkT)^{1/2}\sim 10^{-4}{\rm\ cm},\no
L &=& (\hbar/m\omega)^{1/2}\sim 10^{-4} {\rm\ cm},\no
na^3 &\sim& 10^{-3},
\eea
where $a$ is the S-wave scattering length, $\lambda$  the thermal wavelength,
$L$  the size of the system, and $n$ is the average particle 
density. Thus, $na^3\ll1$, $a/L\ll1$, $a/\lambda\ll1$, and we can treat the 
system as a low-temperature weakly-interacting dilute gas using well-known 
methods [6]. 

In this paper, we review the following theoretical methods, and point out 
some common errors and misconceptions.

(a) The pseudopotential method: One replaces the potential by 
a $\delta$-function, and, with the help of a Bogolubov transformation, 
obtains a weak-coupling expansion. This method seems adequate to describe 
current experiments.

(b) The self-consistent field method: This is variational, and capable of
treating strongly-interacting cases in principle. It
reduces to the pseudopotential method in  appropriate limits; but one should
not use a $\delta$-function potential in a variational calculation, because
such a potential in 3 spatial dimension has no effect, if treated exactly.

(c) The Gaussian variational method: One uses a Gaussian trial wave function, 
and obtains results similar to the self-consistent field method. It has the 
advantage of easily adaptable to the description of time-dependent phenomena.

All three methods are closely related to one another. Although we
will not use Feynman graphs, it might be relevant to note that
the Bogolubov transformation corresponds to 
summing ``one-loop'' graphs. The self-consistent field method consists of 
readjusting parameters in the one-loop sum in a variational sense.
The Gaussian method includes all one-loop contribution, plus parts of
higher-loop contributions. Of these methods, the straightforward one-loop
summation yields a expansion in $na^3$. On the other hand, the 
accuracy of the other methods cannot be ascertained.

We do not attempt a comprehensive review, and the references cited are by 
no means complete. In a subject with such a long 
history, it is difficult to be aware of all relevant sources, and we 
apologize for any omissions [7].

\section{Pseudopotential method}

In low-temperature calculations, one can replace the interatomic
potential with a $\delta$-function potential
$(4\pi a\hbar^2/m)\delta(\r)$, where $a>0$ is the
S-wave scattering length. Such a potential should be treated in
low-order perturbation theory only, because
the $\delta$-function potential
in 3 spatial dimensions has no effect if treated exactly [8].

Consider the Schr\"odinger equation
\be
-\nabla^2\psi(\r)+g\delta^3(\r)=E\psi(\r),
\ee
where $E$ is $2m/\hbar^2$ times the energy. Taking the Fourier transform
of both sides leads to
\be
(E-k^2)\phi(\k)=g\psi(0),
\ee
where $\phi(\k)$ is the Fourier transform of $\psi(\r)$. This must hold for
all $\k$, including $k^2=E$. Therefore $\psi(0)=0$. This condition is
automatically satisfied for all partial waves except S-waves. For S-waves,
this seems to require that $\psi(\r)$ be discontinuous at $\r=0$; but because
of the singular nature of the potential at $\r=0$, one should be careful.
By treating the $\delta$-function
as the limit of a square well, one can verify that in 3 dimensions
the S-waves are the same as those of a free particle except at $\r=0$, 
where they drop to zero discontinuously. This behavior, however, does not 
lead to any scattering, nor level shift. 

As a more careful analysis [9] shows, the $\delta$-function potential must 
be regarded as the first term in a
pseudopotential, which correctly reproduces all scattering phase shifts,
and depends on these phase shifts as input parameters.
For the simple case of a hard-sphere potential, for which there is only one 
parameter $a$, the hard-sphere diameter, the pseudopotential has the form
\be
V_{\rm pseudo}(\r)={4\pi a\hbar^2\over m}\left[\delta^3(\r) 
{\partial\over\partial r}r
-{a^2\over 3}\delta^3(\r)\nabla^2{\partial\over\partial r}r+\cdots
\right],
\ee
The differential operator $(\partial/\partial r)r$ removes spurious 
singularities in the wave function, which would otherwise make the
ground-state energy diverge. In low-order perturbation theory, one can
replace this operator by a simple subtraction procedure for the energy.

Using the first term in the pseudopotential, and making a Bogolubov
transformation, one can calculate exactly the 
first few terms of an expansion for the ground-state energy per particle 
of a uniform hard-sphere Bose gas [10] [11]:
\be
E_0={2\pi n a \hbar^2\over m}\left[1+
{128\over 15}\left(na^3\over \pi\right)^{1/2} + 
8\left({4\pi\over 3}-\sqrt{3}\right)na^3\ln(na^3) + O(na^3)\right].
\ee
This shows that the energy is not analytic at $a=0$, and therefore
cannot be continued to negative values of $a$.

Fetter [12] has given a systematic treatment of the nonuniform case 
based on the Bogolubov transformation. For application to 
the atomic-trap experiments, one can also adapt the results of the uniform case 
using a Thomas-Fermi approximation [5]. 

What happens when the scattering length is negative? In this case, the
two-body scattering is dominated by the attractive part of the potential.
The attraction may or may not be strong enough to produce a two-particle
bound state; but it will lead to an $N$-body bound state,
for a sufficiently large $N$. The many-body problem is very different from 
that of the repulsive case, for it depends on the details of the potential.
Consider two potentials with the same negative scattering length, one
everywhere attractive, and the other having a repulsive core.
The many-body system with the purely attractive potential will collapse
spatially, whereas the one with hard core will yield extensive energies.

When the scattering length is negative, we must therefore use an actual
potential, for example, one with a hard core plus an attractive tail. 
We can replace the hard-core part by a pseudopotential, and do the 
Bogolubov transformation. In this manner, it has been shown [6] [13] that a 
dilute Bose system with such a potential exhibits a first-order phase 
transition separating a gas phase from a liquid phase. The transition line 
terminates in a critical point. The Bose-Einstein condensation occurs in the 
liquid phase, and divides it into liquid I and liquid II. In short, one 
reproduces qualitatively the phase diagram of $^4$He. Since the calculation
is perturbative, one can follow the collapse of the phase diagram to that
of the ideal Bose gas when the potential is turned off.

The results described above pertains to an infinite Bose system in 
the thermodynamic limit, and may be modified for a finite system. 
The authors of Ref.[2] observed Bose condensation in an atomic trap filled 
with in $^7$Li, which has a negative scattering length. In this case,
the smallness of the system may have rendered the first-order transition 
indistinct. It is also possible that the observed phenomenon is metastable.

\section{Variational methods and the energy gap}

A variational approach to the ground-state energy of a 
Bose gas was introduced by Girardeau and Arnowitt [14], who use (wisely) 
an actual potential instead of the $\delta$-function. Their results
are for a uniform gas, but otherwise very 
similar to what we obtain later in the self-consistent field method. 
The excitation spectrum in this approach, however, contains an 
energy gap, contrary to general expectations [15], and to the 
perturbation-theory results in the hard-sphere case [10]. 

A variational method is designed to yield the best ground-state energy,
and may not yield the excitation spectrum accurately. In this instance, 
we know that the gap should have been filled with phonons, which can be
described using general principles. We may therefore view the 
variational principle as a way to obtain an approximate ground-state
wave function, and build the phonon states upon this ground state.

Feynman [16] argues that the low-lying excited states of a Bose
system are purely phonon states, owing to the statistics, and suggests
the following one-phonon wave function of momentum $\k$:
\be
\Psi_\k(\r_1,\ldots, \r_N)=
\sum_{j=1}^N e^{i\k\cdot\r_j}\Psi_0(\r_1,\ldots,\r_N),
\ee
where $\Psi_0$ is the ground-state wave function. The excitation
energy is
\be
E_\k = {\hbar^2k^2\over 2mS_\k},
\ee
where $S_\k$ is the liquid structure factor
\be
S_\k={1\over n}\langle\Psi_0|{\xi_\k}^\dagger\xi_\k|\Psi_0\rangle.,
\ee
where
\be
\xi_k=\Omega^{-1/2}\sum_\p a^\dagger_{\p+\k} a_\p,
\ee
with $a_\k$ the annihilation operator of a free particle of momentum 
$\hbar\k$, and $\Omega$ the volume. From general principles [17] [18], we expect
\be
S_\k\mathop{\longrightarrow}_{\k\to 0} {\hbar k\over 2mc},
\ee
where $c$ is the sound velocity as computed from the
compressibility of the ground state. This leads to the phonon spectrum
\be
E_\k=\hbar ck. 
\ee
Feynman's argument can be sharpened by recasting it as the statement
that the longitudinal sum-rules are saturated
by the phonon states [17] [18]. All the equations above are
verified in the dilute hard-sphere Bose gas [10] [18]. 

Other types of states that can be built upon $\Psi_0$ are those 
describing superfluid flow [10] [18]:
\be
\Psi_{\rm superflow}(\r_1,\ldots,\r_N)=\prod_j^N e^{i\alpha(\r_j)}
\Psi_0(\r_1,\ldots,\r_N),
\ee
where $\alpha(\r)$ is the superfluid phase, which is
related to the superfluid velocity ${\bf v}_s$ through
\be
{\bf v}_s(\r)={\hbar\over m}\nabla\alpha(\r).
\ee
With this wave function one can describe quantized vortices.

\section{Self-consistent field method: ground state}

The self-consistent field method was first used by
Bogolubov in his treatment of superconductivity, and is therefore
also known as the ``Bogolubov method.'' We shall follow
a concise formulation due to De Gennes [19].  A recent review of similar
methods is given by Griffin [20].

We use a quantized-field representation with field operator $\Psi(\x)$, 
and canonical conjugate $i\Psi^\dagger(\x)$:
\be
{[}\Psi(\x),\Psi^\dagger(\y)]=\delta^3(\x-\y).
\ee
The Hamiltonian is
\be
H = \int d^3 x \Psi(\x)^{\dag} h(\x) \Psi(\x)  + \frac{1}{2}\int d^3x d^3y 
\Psi(\y)^{\dag}\Psi(\x)^{\dag} V(\x,\y)
\Psi(\x) \Psi(\y),
\ee
where 
\be
h(\x) = -\frac{\hbar^{2}}{2 m} \nabla^{2} + V_{\rm ext}(\x) - \mu,
\ee
where $V_{\rm ext}(\x)$ is the external potential, 
and $\mu$ is the chemical potential.
We displace the field operator by its ground-state expectation:
\be
\Psi(\x) = \psi(\x) + \phi(\x),
\ee
where $\psi(\x)$ is the displaced field operator, and 
\be
\phi(\x) =\langle\Psi(\x) \rangle.
\ee
The function $\phi(\x)$ describes a nonuniform condensate, and $\psi(\x)$
annihilates particles not in the condensate.

Variational trial states are taken to be the eigenstates of
an effective Hamiltonian that is quadratic in the field operators, chosen
to be a mean-field version of the actual Hamiltonian. There are two 
variational functions $\rho(\x,\y)$  and $\Delta(\x,\y)$,
which will turn out to be 
$\rho(\x,\y)=\langle \psi^\dagger(\x)\psi(\y)\rangle$, and  
$\Delta(\x,\y)=\langle \psi(\x)\psi(\y)\rangle$. That is, $\rho$ and $\Delta$
are, respectively, the direct and off-diagonal density correlation functions
of the uncondensed particles.

The effective Hamiltonian is chosen to be
\be
H_{\rm eff} = H^{(0)}+ H^{(1)}+H^{(1)\dagger}  + H^{(2)},
\ee
where $H^{(0)}$ is independent of $\psi$, and $H^{(1)}$ and $H^{(2)}$
terms are respectively linear and quadratic in $\psi$:
\be
H^{(0)} = \int d ^3x \phi^\ast(\x) h(\x) \phi(\x) +
{1\over 2}\int d^3x d^3y \phi^\ast(\y) \phi^\ast(\x) V(\x,\y) \phi(\x)\phi(\y),
\ee
\bea
H^{(1)} &=& \int d^3x \psi^\dagger(\x) h(\x) \phi(\x) \no
&& + \int d^3x d^3y V(\x,\y) \psi^\dagger(\x)  \left[\phi(\y) \rho(\y,\x) 
 + \phi(\x) \rho (\y,\y) + \phi^\ast(\y) D(\x,\y) \right],
\eea
\bea
H^{(2)} &=& \int d^3x \psi^\dagger(\x) h(\x) \psi(\x) \no
 &&+  \int d^3x d^3y V(\x,\y)\left[ \psi^\dagger(\y) \psi(\x) R(\x,\y)
+ \psi^\dagger(\x) \psi(\x) R(\y,\y) \right]\no
&&+\frac{1}{2}\int d^3x d^3y V(\x,\y) \left[ \psi^\dagger(\x) \psi^\dagger(\y)
 D(\x,\y) + \psi(\x)\psi(\y) D^\ast(\x,\y)\right],
\eea
where
\bea
R(\x,\y) &\equiv& \rho(\x,\y) + \phi^\ast(\x) \phi(\y),\no
D(\x,\y) &\equiv& \Delta(\x,\y) + \phi(\x) \phi(\y).
\eea

For arbitrary $\rho$ and $\Delta$, we can diagonalize the effective
Hamiltonian through a generalized Bogolubov transformation:
\be
\psi(\x) = \sum_{n}\left[ u_{n}(\x)\eta_{n} 
- v_{n}^{\ast}(\x) \eta_{n}^{\dag} \right].
\ee
where $\eta_n$ and $\eta^\dagger_n$ are annihilation and creation operators 
satisfying 
\be
[\eta_n,\eta^\dagger_m]=\delta_{nm}.
\ee
To preserve the canonical commutators for $\psi(\x)$, we require
\be
\sum_{n} \left[ u_{n}(\x)u^\ast_{n}(\y) - v_{n}^\ast(\x)
v_{n}(\y) \right] = \delta^3(\x - \y).
\label{e:condition}
\ee
We now require $H_{\rm eff}$ to have the form
\be
H_{\rm eff} = E_0 + \sum_{n} \epsilon_{n} \eta_{n}^{\dag} \eta_{n},
\ee
or, equivalently,
\bea
{[} H_{\rm eff},\eta_{n}] &=& - \epsilon_{n} \eta_{n}, \no
{[} H_{\rm eff},\eta_{n}^{\dag}] &=& \epsilon_{n} \eta_{n}^{\dag}.
\eea
These conditions determine the functions $v_{n}$, $u_{n}$ and $\phi$. 
The condition for $\phi$ comes from the requirement that in 
$H_{\rm eff}$ the terms linear in $\psi(\x)$ vanish.
The resulting equations are
\bea
h(\x) \phi(\x) &+& \int d^3 y V(\x,\y) \left[\phi(\y)  \rho(\y,\x)
+\phi(\x)\rho(\y,\y) \right. \no
&&+ \left. \phi^\ast(\y) \Delta(\x,\y) + |\phi(\y)|^{2} \phi(\x) \right] =0,
\label{e:NLSE}
\eea
\be
\epsilon_{n}u_{n}(\x) = h(\x) u_{n}(\x) +  
\int d^3y V(\x,\y) \left[ u_{n}(\x) R(\y,\y)
 + u_{n}(\y) R(\y,\x) -  v_{n}(\y) D(\x,\y) \right],
\label{e:uequation}
\ee
\be
-\epsilon_{n} v_{n}(\x) = h(\x) v_{n}(\x) + 
\int d^3y V(\x,\y) \left[ v_{n}(\x) R(\y,\y)
+ v_{n}(\y) R(\y,\x) - u_{n}(\y) D^\ast(\x,\y) \right].
\label{e:vequation}
\ee

To determine $\rho$ and $\Delta$, let $|\Phi\rangle$ be the lowest eigenstate
of $H_{\rm eff}$:
\be
H_{\rm eff} | \Phi \rangle = E_0 | \Phi \rangle.
\ee
We require that the ground-state energy be at a minimum with respect to
variations in $\rho$ and $\Delta$:
\be
\delta \langle H \rangle = 0,
\ee
where the brackets denote expectation value with respect to $|\Phi\rangle$.
In calculating the expectation value above, we use Wick's theorem: 
\bea
\langle \psi^\dagger(\x) \psi^\dagger(\y) \psi(\x) \psi(\y)
\rangle &=&  \langle \psi^\dagger(\x)\psi(\y)\rangle \langle
\psi^\dagger(\y) \psi(\x) \rangle +
 \langle \psi^\dagger(\x) \psi(\x)\rangle\langle\psi^\dagger(\y)
\psi(\y)\rangle \no
& & +\langle \psi^\dagger(\x) \psi^\dagger(\y)\rangle\langle
\psi(\x) \psi(\y) \rangle.
\eea
The calculation is facilitated by noting that 
$\delta\langle H_{\rm eff}\rangle=0$, since $\Phi$ is an eigenstate. Thus
we can use the equivalent condition
\be
\delta \langle H - H_{\rm eff} \rangle = 0.
\ee
The results are as follows:
\bea
\rho(\x,\y) &=& \langle \psi^\dagger(\x) \psi(\y)\rangle
=\sum_{n} v_{n}(\x) v_{n}^\ast(\y),\no
\Delta(\x,\y) &=& \langle \psi(\x) \psi(\y)\rangle
= -\sum_{n} u_{n}(\x) v_{n}^\ast(\y).
\label{e:rhodelta}
\eea
When these are substituted into (\ref{e:NLSE}), ({\ref{e:uequation}),
and (\ref{e:vequation}), we have a system of coupled nonlinear 
integro-differential equations.

\section{The uniform case}
We put $V_{\rm ext}=0$, assume $V(\x,\y)=V(\x-\y)$, and
seek solutions with uniform density by putting
\bea
\phi(\r)   &=& \phi_{0},\no
u_{\k}(\r) &=& \Omega^{-1/2}e^{i \k\cdot\r} x_{k},\no
v_{\k}(\r) &=& \Omega^{-1/2} e^{i \k\cdot\r} y_{k},
\eea
where $x_{k}$, $y_{k}$ and $\phi_{0}$ can be taken to be real. The condition
(\ref{e:condition}) becomes $x_{k}^{2} - y_{k}^{2} =1$, which can be
satisfied by putting
\bea
x_{k} &=& \cosh \sigma_{k},\no
y_{k} &=& \sinh \sigma_{k}.
\eea
This parametrization simplifies the calculations. For example,
the liquid-structure factor of the ground state can be represented in 
the form
\bea
S_\k &=& 1+{n_0\over n} (\cosh2 \sigma_{k} -\sinh 2 \sigma_{k} -1) \no
&& +{1\over 4n\Omega} 
\sum_{\p} \left[ \sinh 2 \sigma_{p} \sinh 2 \sigma_{|\k + \p|}
 + \sinh^{2} \sigma_{p} \sinh^{2} \sigma_{|\p+\k|} \right],
\eea
where $n_0$ is the condensate density given through
\be
n=n_0+\Omega^{-1}\sum_\p \sinh^2\sigma_\p.
\ee
The Bogolubov result is recovered by neglecting the $\p$ sums,
which is equivalent to putting $n_0\approx n$:
\be
S_\k=\left[ \epsilon_k\over \epsilon_k +2nV_k\right]^{1/2},
\ee
where $\epsilon_k=\hbar^2k^2/2m$, and $V_k$ is the Fourier transform of 
the interaction potential.

\section{Reduction to $\delta$-function potential}

The equations in the last section are rather complex, and difficult to
solve even numerically. To gain some insight, we reduce them to a simpler
form by using a $\delta$-function potential. We shall show that this 
potential has no effect if self-consistency is enforced, at least
in the case of a uniform system. We put
\be
V(\x,\y)=g\delta^3(\x-\y), 
\ee
where
\be
g=4\pi\hbar^2 a/m.
\ee
The functions $\rho$ and $\Delta$ now depend only on $\x$:
\bea
\rho(\x) &=& \langle \psi^\dagger(\x) \psi(\x)\rangle
=\sum_{n} v_{n}(\x) v_{n}^\ast(\x)\no
\Delta(\x) &=& \langle \psi(\x) \psi(\x)\rangle
= -\sum_{n} u_{n}(\x) v_{n}^\ast(\x).
\label{e:selfcon}
\eea
Note that $\rho$ is the depletion of the condensate due to interactions.
The equations (\ref{e:NLSE}), ({\ref{e:uequation}),
and (\ref{e:vequation}) reduce to
\be
\left[h +2 g\rho + g|\phi|^2\right]\phi +g\Delta\phi^\ast =0,
\label{e:NL}
\ee
\bea
(h+2g\rho+2g\phi^\ast\phi)u_n -g(\Delta+\phi^2)v_n &=& \epsilon u_n,\no
(h+2g\rho+2g\phi^\ast\phi)v_n -g(\Delta^\ast +
\phi^{\ast 2})u_n &=& -\epsilon v_n,
\label{e:uv}
\eea
where we have suppressed the $\x$ dependence of the functions.
The dilute uniform hard-sphere gas is recovered by setting $\rho=\Delta=0$,
for they give higher-order contributions.

For $\rho=\Delta=0$, (\ref{e:NL}) is the 
``nonlinear Schr\"odinger equation'' [21] [22], which has been widely used
in numerical studies of the condensate and its excitations [23]. One 
should be aware, however, that $\rho$ and $\Delta$ may be important,
especially for the excitations.

In the uniform case, with $V_{\rm ext}=0$, we obtain, in the notation
of the last section,
\be
\tanh 2\sigma_k = { g(n-\rho+\Delta)\over \epsilon_k +g(n-\rho-\Delta)},
\ee
where $\epsilon_k=\hbar^2k^2/2m$, and the chemical potential $\mu$ has 
been eliminated in terms of the density $n$.
The self-consistency conditions (\ref{e:selfcon}) then lead to 
\bea
\rho &=& {1\over 4\pi^2}\int_0^\Lambda dk k^2 \left\{
f_k \left[\epsilon_k + g(n-\rho-\Delta)\right]-1\right\}, \no
\Delta &=& -{g\over 4\pi^2}\int_0^\Lambda dk k^2 f_k
(n-\rho+\Delta),
\eea
where
\be
f_k =\left[ \epsilon_k^2+2g\epsilon_k(n-\rho-\Delta)
-4g^2(n-\rho)\Delta\right]^{-1/2}.
\ee
The integrals are cut off at $\Lambda$, because otherwise
$\Delta$ would be divergent. Solving for $\Delta$ from the second equation, 
we obtain in the limit $\Lambda\to\infty$
\be
\Delta-\rho+n=0.
\ee
The first equation then gives, in the limit $\Lambda\to\infty$,
\be
\rho=0.
\ee
That is, there is no depletion of the condensate. It is then straightforward
to show that the system is an ideal Bose gas. This result is reassuring;
it shows that the method is able to verify that the $\delta$-function
potential has no effect.

\section{Self-consistent field method: finite temperatures}

To extend the variational approach to finite temperatures, we can
use the eigenstates of $H_{\rm eff}$ to calculate the
partition function, and then minimize the free energy
\be
F=\langle H \rangle -TS,
\ee
where the brackets now denote thermal average with respect to $H_{\rm eff}$,
and $S$ is the entropy. De Gennes [19] argues that the quantity
\be
F_{\rm eff}=\langle H_{\rm eff}\rangle -TS
\ee
is stationary, and implements the variational principle
by minimizing $F-F_{\rm eff}$. But the assertion is
correct only if $S$ were defined with respect to the eigenvalue spectrum 
of $H_{\rm eff}$. In fact, as we explain below, $S$ is defined with respect to
the energy levels $\langle \alpha|H|\alpha \rangle$, where $|\alpha \rangle$  
are the eigenstates of $H_{\rm eff}$. The finite-temperature results in Ref.[19]
are therefore incorrect.

A proper variational principle for the free energy [24] is based on the
inequality
\be
\sum_\alpha \langle \alpha| e^{-\beta H}|\alpha \rangle 
\ge \sum_\alpha e^{-\beta \langle \alpha| H |\alpha\rangle}.
\ee
This shows that, to calculate the trial partition function, we must 
use the energy spectrum
\be
W_\alpha = \langle \alpha| H |\alpha\rangle.
\label{e:correct}
\ee
In contrast, the other scheme leads to a free Bose gas 
with single-particle energy $\epsilon_n$, the eigenvalues 
from (\ref{e:uequation}) and (\ref{e:vequation}). Because of the simplicity,
it is widely used in the literature; but it
does not follow from a variational principle.

As an example of the difference between the two schemes, the calculation
of the partition of a hard-sphere Bose gas in Ref.[25] corresponds to 
using (\ref{e:correct}), and yields the virial coefficients in the gas phase. 
On the other hand, a calculation based on the other scheme would give
a free Bose gas above the transition temperature.

\section{Gaussian variational method}

We go to a representation in which the field operator $\Psi(\x)$ and its
conjugate have the forms
\bea
\Psi(\x) &=& \frac{1}{\sqrt{2}} \left[\varphi(\x) 
+ {\delta\over\delta\varphi(\x)}\right], \no
i\Psi^\dagger(\x) &=& \frac{i}{\sqrt{2}} \left[\varphi(\x) 
- {\delta\over\delta\varphi(\x)}\right],
\eea
where $\varphi(\x)$ is a c-number function. The more familiar representation,
in which $\Psi$ is diagonal, would be awkward to use in a non-relativistic
case, in which $\Psi$ and $i\Psi^\dagger$ are canoncial conjugates.
(A similar situation happens in a relativistic fermion field obeying
the Dirac equation.)

The state of the system is described by a wave functional $\Phi[\varphi]$.
We use a Gaussian form for a trial wave functional:
\be
\Phi[\varphi] = C \exp\left\{-{1\over 4}\int d^3x d^3y [\varphi(\x) -
\varphi_{0}(\x)] G^{-1}(\x,\y)[\varphi(\y) -
\varphi_{0}(\y)]\right\},
\ee
where $C$ is a normalization constant and $\varphi_{0}$ and 
$G(\x,\y)$ are the variational parameters.  
 Expectation values are given by functional integrals, for example 
\be
\langle H \rangle = \int ( D \varphi)
\Phi^{\ast}[\varphi] H \Phi[\varphi]
\ee
where $\Phi$ is normalized to unity. It is easy to show that
\bea
\langle \Psi(\x)\rangle &=& \varphi_0(\x),\no
\langle \varphi(\x)\varphi(\y)\rangle &=& G(\x,\y) +\varphi_0(\x)\varphi_0(\y).
\eea
The variational principle states that
\be
{\delta \langle H\rangle\over \delta G(\x,\y)}=0,\qquad
{\delta \langle H\rangle\over \delta\varphi_0(\x)}=0.
\ee

The equation obtained for $G$ and $\varphi_0$ will not be given 
directly. To make contact with 
the self-consistent field method, we quote the results
\bea
\rho(\x,\y) &\equiv& \langle \psi(\x)^{\dag} \psi(\y)\rangle = 
\frac{1}{2} \left[{1\over 4}G^{-1}(\x,\y) + G(\x,\y)
-\delta^3(\x-\y)\right],\no
\Delta(\x,\y) &\equiv& \langle \psi(\x) \psi(\y)\rangle = 
-\frac{1}{2} \left[{1\over 4} G^{-1}(\x,\y) -G(\x,\y) \right].
\eea
For further analysis, it is convenient to introduce
\bea
X(\x,\y) &\equiv& \frac{1}{2 \sqrt{2}} \left[ G^{-1/2}(\x,\y) 
+ 2 G^{1/2} (\x,\y) \right],\no
Y(\x,\y) &\equiv& \frac{1}{2 \sqrt{2}} \left[ G^{-1/2}(\x,\y)
-2 G^{1/2} (\x,\y) \right].
\eea
it easy to show that $\rho(\x,\y) = \int d^3z Y(\x,\z) Y(\z,\y)$, and
$\Delta(\x,\y) = -\int d^3z X (\x,\z) Y(\z,\y)$. Now expand $X$ and $Y$ 
in terms of a basis $\{b_n(\x)\}$:
\bea
X(\x,\y)=\sum_n X_n b_n(\x)b_n^\ast(\y), \no
Y(\x,\y)=\sum_n Y_n b_n(\x)b_n^\ast(\y),
\eea
where $X_n$ and $Y_n$ are real. We then find
\bea
\rho(\x,\y) &=& \sum_{n} b_{n}(\x) b_{n}^\ast(\y) Y_n^2, \no
\Delta(\x,\y) &=& -\sum_{n} b_{n}(\x) b_{n}^\ast(\y) X_n^2.
\eea
Comparison with (\ref{e:rhodelta}) leads to the identification
\bea
u_{n}(\x) &=& b_{n}(\x) X_{n},\no
v_{n}(\x) &=& b_{n}(\x) Y_{n}.
\eea
This shows the equivalence between this method and the self-consistent
field method.

One of the advantages of the Gaussian method lies in the possibility of
generalization to the ``time-dependent Hartree-Fock'' method [26] [27], 
which has been used successfully in nuclear physics. One of us (P.T.) plans to
discuss this in a separate publication.

\section*{Acknowledgments}
We thank R. Jackiw for suggesting the field representation used
in the last section. This work was supported in part by funds provided by 
the U.S. Department of Energy under cooperative agreement 
\# DE-FC02-94ER40818.
P.T. was supported by Conselho Nacional Desenvolvimento Cientifico e
Tecnologico (CNPq), Brazil.
\newpage

\section*{References}
\begin{enumerate}
\item M.H. Anderson, J.R. Ensher, M.R. Mathews, C.E. Wieman, and E.A. Cornell,
Science, {\bf 269}, 198 (1995).
\item C.C. Bradley, C.A. Sackett, J.J. Tollett, 
and R.G Hulet, Phys. Rev. Lett. {\bf 75}, 1687 (1955)
\item  K.B. Davis, M.O. Mewes, M.R. Andrews, N.J. van Druten, D.D. Durfee, 
D.M. Kum, and W. Ketterle, Phys. Rev. Lett. {\bf 75}, 3969 (1955).
\item G. Baym and C. Pethick, ``Ground-state properties of magnetically
trapped Bose-condensed rubidium gas,'' University of Illinois preprint (1995).
\item T.T. Chou, C.N. Yang, and L.H. Yu, ``Bose-Einstein condensation of atoms
in a trap,'' Stony Brook preprint (1996). 
\item For a review see K. Huang in {\it Studies in Statistical Mechanics}, Vol. II, 
J. De Boer and G.E. Uhlenbeck, Eds. (North-Holland, Amsterdam, 1964), Chap.1.
\item For other references see {\it Bose-Einstein Condensation}, A. Griffin, 
D.W. Snoke, and S. Stringari, Eds. (Cambridge University Press, 
Cambridge, England, 1995).
\item K. Huang, J. Mod. Phys. (Singapore) {\bf 4}, 1037 (1989).
\item K. Huang and C.N. Yang, Phys. Rev. {\bf 105} 767 (1957).
\item T.D. Lee, K. Huang, and C.N. Yang, Phys. Rev. {\bf 106}, 1135 (1957).
\item T.T. Wu, Phys. Rev. {\bf 115}, 1390 (1959).
\item A.L. Fetter, Ann. Phys. (NY) {\bf 70}, 67 (1972).
\item K. Huang, Phys. Rev. {\bf 115}, 769 (1959); 
Phys. Rev. {\bf 119}, 1129 (1960).
\item M. Girardeau and R. Arnowitt, Phys. Rev. {\bf 113}, 755 (1959).
\item N.M. Hugenholtz and D. Pines, Phys. Rev. {\bf 116}, 489 (1959).
\item R.P. Feynman, Phys. Rev, {\bf 94}, 262 (1954).
\item K. Huang and A. Klein, Ann. Phys. (NY) {\bf 30}, 203 (1964).
\item K. Huang, {\it Statistical Mechanics}, 2nd ed. (Wiley, New York, 1987), 
Chap.13.
\item P.G. De Gennes, {\it Superconductivity of Metal and Alloys}
(Benjamin, New York, 1966), Chap.5. Note that
the finite-temperature treatment is incorrect. (See Sec.7 of this paper.)
\item A. Griffin, ``Conserving and gapless approximations for an inhomogeneous
Bose gas at finite temperatures,'' University of Toronto preprint (1995). 
\item E.P. Gross, N. Cimento, {\bf 20}, 454 (1961); 
J. Math. Phys. {\bf 4}, 195 (1963).
\item V.L. Ginzburg and L.P. Pitaevskii, Sov. Phys. JETP {\bf 7}, 858 (1958).
\item P.A. Ruprecht, M.J. Holland, K. Burnett, and M. Edwards, Phys. Rev.
A{\bf 51}, 4704 (1995).
\item  K. Huang, {\it op. cit.}, Sec. 10.4.
\item K. Huang, C.N. Yang, and J.M. Luttinger, 
Phys. Rev. {\bf 105}, 776 (1957).
\item A.K. Kerman and S.E. Koonin, Phys. Rev. D{\bf 40}, 504 (1989).
\item J.W. Negele, Rev. Mod. Phys. {\bf 54}, 913 (1986).
\end{enumerate}  
\end{document}